 \documentclass[preprint2]{aastex}

\shorttitle{Zhang \& Cheng}
\shortauthors{
$\gamma$-ray Luminosity and Death Lines of Pulsars with Outer Gaps}

\begin{document}

\title{Gamma-ray Luminosity and Death Lines of Pulsars with Outer Gaps}

\author{L. Zhang$^{1,2}$, K.S. Cheng$^3$, Z.J. Jiang$^2$, P. Leung$^3$}
\affil{$^1$National Astronomical Observatories/Yunnan Observatory,
Chinese Academy of Sciences, P.O. Box 110, Kunming, PRC\\
$^2$Department of Physics, Yunnan University, Kunming,
PRC\\
$^3$Department of Physics, the University of Hong Kong, Hong Kong,
PRC }

\begin{abstract}
We re-examine the outer gap size by taking  the geometry of dipole
magnetic field into account. Furthermore, we also consider that
instead of taking the gap size at half of the light cylinder
radius to represent the entire outer gap it is more appropriate to
average the entire outer gap size over the distance. When these
two factors are considered, the derived outer gap size $f(P, B,
<r>(\alpha))$ is not only the function of period ($P$) and
magnetic field ($B$) of the neutron star, but also the function of
the average radial distance to the neutron star $<r>$; which
depends on the magnetic inclination angle ($\alpha$). We use this
new outer gap model to study $\gamma$-ray luminosity of pulsars,
which is given by $L_{\gamma} = f^3(P, B, <r>(\alpha))L_{sd}$ and
$L_{sd}$ is the pulsar spin-down power, as well as the death lines
of $\gamma$-ray emission of the pulsars. Our model can predict the
$\gamma$-ray luminosity of individual pulsar if its $P, B$ and
$\alpha$ are known. Since different pulsars have different
$\alpha$, this explains why some $\gamma$-ray pulsars have very
similar $P$ and $B$ but have very different $\gamma$-ray
luminosities. In determining the death line of $\gamma$-ray
pulsars, we have used a new criterion based on concrete physical
reason, i.e. the fractional size of outer gap at the null charge
surface for a given pulsar cannot be larger than unity. In
estimate of the fractional size of the outer gap, two possible
X-ray fields are considered: (i) X-rays are produced by the
neutron star cooling and polar cap heating, and (ii)X-rays are
produced by the bombardment of the relativistic particles from the
outer gap on the stellar surface (the outer gap is called as a
self-sustained outer gap). Since it is very difficult to measure
$\alpha$ in general, we use a Monte Carlo method to simulate the
properties of $\gamma$-ray pulsars in our galaxy. We find that
this new outer gap model predicts many more weak $\gamma$-ray
pulsars, which have typical age between 0.3-3 million years old,
than the old model. For all simulated $\gamma$-ray pulsars with
self-sustained outer gaps, $\gamma$-ray luminosity $L_{\gamma}$
satisfies $L_{\gamma}\propto L^{\delta}_{sd}$; where the value of
$\delta$ depends on the sensitivity of the $\gamma$-ray detector.
For the EGRET, $\delta$ is $\sim 0.38$ whereas $\delta$ is $\sim
0.46$ for the GLAST. For $\gamma$-ray pulsars with $L_{sd}
\lesssim L_{sd}^{crit}$, $\delta$ is  $\sim 1$. $L^{crit}_{sd} =
1.5 \times 10^{34} P^{1/3} ergs^{-1}$ is determined by $f(<r>\sim
r_L) =1$. These results are roughly consistent with the observed
luminosity of $\gamma$-ray pulsars. These predictions are very
different from those predicted by previous outer gap model, which
predicts a very flat relation between $L_{\gamma}$ and $L_{sd}$.
\end{abstract}

\section{Introduction}

High-energy emission models for rotation-powered pulsars are
generally divided into polar gap and outer gap models. In polar
gap models, charged particles are accelerated in charge-depleted
zones near the pulsar's polar cap and $\gamma$-rays are produced
through curvature-radiation induced $\gamma$-B pair cascade (e.g.
Harding 1981; Daugherty \& Harding 1996; Zhang \& Harding 2000) or
through Compton-induced pair cascades \cite{dermer94}. In the
outer gap models, it is generally accepted that a magnetosphere of
charge density
\begin{equation}
\rho_0\approx {{\bf{\Omega}}\cdot{\bf{B}}\over 2\pi c}
\end{equation}
surrounds a rotating neutron star with magnetic field $B$ and
angular velocity $\Omega$ (Goldreich \& Julian 1969). The
magnetospheric plasma is corotating with the neutron star within
the light cylinder, at which the corotating speed equals the
velocity of light and the distance from the spin axis is

$R_L=c/\Omega$. In the corotating magnetosphere, the electric
field along the magnetic field, $E_{||}={\bf{E}}\cdot{\bf{B}}/B$,
is nearly zero. However, the flows of the plasma along open field
lines will results in some plasma void regions (where the charge
density is different significantly from $\rho_0$) in the vicinity
of null charge surfaces where ${\bf{\Omega}}\cdot{\bf{B}}=0$
(Holloway 1973). In such charge deficient regions, which are
called outer gaps, $E_{||}\neq 0$ is sustained,
electrons/positrons can be accelerated to relativistic energies
and the subsequent high-energy gamma-ray emission and
photon-photon pair production can maintain the current flow in the
magnetosphere (Cheng, Ruderman \& Sutherland 1976; Cheng, Ho \&
Ruderman 1986a, 1986b, hereafters CHRI and CHR II; Romani, 1996;
Zhang \& Cheng 1997; Hirotani 2001 ).

Based on known $\gamma$-ray pulsars, the luminosity and conversion
efficiency of $\gamma$-rays in various models have been studied
(for example, Harding 1981; Dermer \& Sturner 1994; Rudak \& Dyks
1998; Yadigaroglu \& Romani 1995; Zhang \& Cheng 1998).
Observations by Compton Gamma-ray Observatory (CGRO) show that
$\gamma$-ray luminosity of rotation-powered pulsars is
proportional to square root of the spin-down power (Thompson et
al. 2001). Using the recent new polar gap models (e.g. Zhang \&
Harding 2000; Harding \& Muslimov 2001; Harding, Muslimov \&
Zhang, 2002), Harding et al. (2002) have studied the deadline of
$\gamma$-ray pulsars based on the predicted luminosity of
$\gamma$-ray pulsars $L_{\gamma}\propto L^{\delta}_{sd}$, where
$\delta \sim 0.5$ when $L_{sd} \gtrsim L^{break}_{sd}$ and $\delta
\sim 1.$ when $L_{sd} \lesssim L^{break}_{sd}$ respectively and
$L^{break}_{sd} = 5 \times 10^{33} P^{-1/2}$ erg/s.

For a rapid rotating pulsar, it is believed that its spin-down
power, $L_{sd}$, is converted into radiation energy. Because the
outer gap occupies only a part of the open field line region, the
gamma-ray luminosity produced in the outer gap is a fraction of
the spin-down power. It has been shown that the gamma-ray
luminosity in the outer gap is proportional to $f^3$, {\bf{i.e
$L_{\gamma}\approx f^3L_{sd}$}} (CHR II; Zhang \& Cheng 1997). In
previous works, the factional size of outer gap only depends on
the period and magnetic field for a gamma-ray pulsar. For example,
the fractional size of the outer gap for Crab-like pulsars is
$f\propto B_{12}^{-13/20}P^{33/20}$ (CHR II). In the outer gap
model described by Zhang \& Cheng (1997), $f\propto
B^{-4/7}P^{26/21}$, the observed gamma-ray luminosities with
energies greater than 100 MeV from the known gamma-ray pulsars
except for the Crab pulsar can be explained approximately (Zhang
\& Cheng 1998). It should be noted that the model predicts that
the $\gamma$-ray pulsars with the same values of $(B/P)^{0.3}$
have same $\gamma$-ray luminosities. For example, the ratio of
$(B/P)^{0.3}$ for PSR B1055-52 to that for Geminga is $\sim 0.9$,
it means that the ratio of both gamma-ray luminosities is $\sim
0.9$. However, the observed ratio of the luminosities with
energies greater than 100 MeV of these two pulsars are $\sim 8$
(Kaspi et al. 2000). Briefly, 7 known $\gamma$-ray pulsars have
rather different $\gamma$-ray luminosity even their spin-down
powers and ages are so similar (e.g. Geminga and PSR 1055-52).
Their pulse shapes also differ so much. It is clear that there are
other intrinsic parameters to control these observed properties
(gamma-ray luminosity, pulse shape , spectrum etc).

 In this paper, we re-study the gamma-ray emission from the
outer gaps of the rotation-powered pulsars by using a new outer
gap model. We follow the idea of self-sustained outer gap given by
Zhang \& Cheng (1997). However, we take the magnetosphere geometry
as well as the average properties of the entire outer gap into
consideration and show that the fractional size of the outer gap
is a function of period, magnetic field and magnetic inclination
angle. In fact, the effect of the inclination angle on the
$\gamma$-ray emission have been considered in other versions of
the outer gap models. For example, Romani \& Yadigaroglu (1995)
and Yadigaroglu \& Romani (1995) took the magnetic inclination
angle into account in their outer gap models, Hirotani and his
colleagues also included the magnetic inclination angle in their
calculation of the outer gap model (e.g. Hirotani, 2001; Hirotani
\& Shibata 2001; Hirotani \& Shibata 2002: Hirotani, Harding,
Shibata 2003). Different from the treatment of Romani \&
Yadigaroglu (1995) who considered it in a less analytic way, we
give a explicit expression for the fractional size of the outer
gap.  In section 2, we describe the revised outer gap model. We
estimate the $\gamma$-ray luminosity for rotation-powered pulsars
and compare them with the observed data in section 3. In section
4, we derive the death lines of the pulsars with outer gaps.
Finally, we give briefly our conclusion and discussion.

\section{The Outer Gap Model}

\subsection{Magnetospheric Geometry}

For an oblique magnetic dipole rotator with an angular velocity
${\bf{\Omega}}$ and the magnetic moment vector ${\bf{\mu}}$, let its spin
axis be along the Oz axis, ${\bf{\mu}}$ be in the plane xOz and $\alpha$
be the angle between ${\bf{\Omega}}$ and ${\bf{\mu}}$. In polar
coordinates,
\begin{equation}
{\bf{\Omega}}=\Omega (\cos\theta \hat{r}- \sin\theta \hat{\theta})\;\;.
\label{Omegav}
\end{equation}
and ${\bf{\mu}}=\mu(\cos(\theta-\alpha)\hat{r},
-\sin(\theta-\alpha)\hat{\theta})$. Corresponding magnetic field is
\begin{equation}
\bf{B(r)}={\mu\over 2r^3}(2\cos(\theta-\alpha)\hat{r}+
\sin(\theta-\alpha)\hat{\theta})\;\;\;,
\label{Bvector}
\end{equation}
where $\mu=B_pR^3/2$, $B_p$ and $R$ are the stellar radius and surface
magnetic field at pole (see, for example, Zhang \& Harding 2000).
$\hat{r}$ and $\hat{\theta}$ are the unit vectors of radial
and polar angle directions respectively. For a pulsar with a period $P$
and a period derivative $\dot{P}$, the $B_p$ is estimated by
\begin{equation}
B_p\approx 6.4\times 10^{19}(P\dot{P})^{1/2}\;\; {\mbox{G}}\;\;.
\label{Bp}
\end{equation}

It is believed that the outer gap is extended from its inner boundary to
the light cylinder (CHR I). For the oblique magnetic dipole rotator, the
polar angle, $\theta_c$,  in which the last open field line is tangent to
the light cylinder given by (Kapoor \& Shukre 1998)
\begin{equation}
\tan\theta_c=-{3\over 4\tan\alpha}(1+(1+8\tan^2\alpha/9)^{1/2})\;\;,
\label{thetac}
\end{equation}
corresponding radius is
\begin{equation}
r_c={R_L\over \sin\theta_c}\;\;,
\label{rc}
\end{equation}
it should be noted that $\theta_c=\pi/2$ and $r_c=R_L$ for an aligned
magnetic dipole. Along the last open field line, the relation
\begin{equation}
{\sin^2(\theta-\alpha)\over r}={\sin^2(\theta_c-\alpha)\over r_c}
\label{sinlaw}
\end{equation}
is valid. The inner boundary of the outer gap is estimated by the null
charge surface, which is defined by ${\bf{\Omega}}\cdot{\bf{B}}=0$. In the
two dimensional case, the null charge surface can be described by
($r_{in}$, $\theta_{in}$). By definition, we have
\begin{equation}
\tan\theta_{in}={1\over 2}(3\tan\alpha+\sqrt{9\tan^2\alpha+8})\;\;\;,
\label{thetain}
\end{equation}
while $r_{in}$ is estimated along the last open field line, which gives
using Eq. (\ref{sinlaw})
\begin{equation}
{r_{in}\over R_L}={\sin^2(\theta_{in}-\alpha)\over \sin\theta_c
\sin^2(\theta_c-\alpha)}\;\;\;.
\label{rin}
\end{equation}
 Therefore, the outer gap extended from $r_{in}$ to  $r_c$ along the last
open field line for the oblique magnetic dipole. In such a geometry, the
Goldreich-Julian current is roughly
\begin{equation}
\dot{N}_{GJ}\approx {\Omega^2R^3B_p\over 2 ec}
a(\alpha)\cos\alpha\;\;, \label{NGJ}
\end{equation}
where $a(\alpha)=\sin\theta_c \sin^2(\theta_c-\alpha)$.

\subsection{X-ray Field in the Magnetosphere}

The observed spectra of X-ray emission from some rotation-powered
pulsars indicate that there are at least two kinds of X-ray
spectra. One consists of soft X-rays, which can be fitted by a
black-body spectrum with a single temperature, combined with a
hard X-ray spectrum with a power-law distribution. Another has
only a thermal spectrum. It is believed that the non-thermal
components are most likely of magnetospheric origin, while the
origin of thermal components is less clear because there are
several mechanisms for production of thermal emission in the soft
X-ray bands. Observationally, the bulk of the soft X-ray emission
between $\sim 0.1$- 1 keV is well fitted by a double black-body at
two different temperatures. The ratio of the area of the hotter
component to the colder one is typically very small ($\sim$ a few
$\times 10^{-5}$ to a few $\times 10^{-5}$). The colder component
has been explained as resulting from thermal cooling, while the
hotter one most likely comes from bombardment by high-energy
particles (Greiveldinger et al. 1996). It is expected that pulsars
younger than $\sim 10^5$ yr have significant neutron star cooling
components. We describe briefly the main possible mechanisms of
X-ray production as below.

\subsubsection{Thermal X-ray emission from neutron star cooling}

For thermal X-ray emission due to neutron star cooling, it is
believed that its spectrum is expressed with a modified
black-body, we approximate it as a black-body spectrum with a
temperature $T_c$. Because the neutron star cooling concerns many
different mechanisms, the estimate of the temperature $T_c$ exists
uncertainties for different models. For example, the temperature
can be expressed as (Romani 1996)
\begin{equation}
T_{c,6}\approx \left({\tau\over 10^5{\mbox{yr}}}\right)^{-0.05}
\exp(-\tau/10^6{\mbox{yr}})\;\;\; {\mbox{K}}\;\;,
\end{equation}
where $T_{c,6}$ is the temperature of the neutron star cooling in
units of $10^6$ K and $\tau$ is the neutron star's age. Above
expression is valid for $\tau$ less than several $10^6$ yr. Zhang
\& Harding (2000) used a expression to approximate the temperature
$T_c$ as
\begin{equation}
T_{c,6}=\left\{ \begin{array}{ll}
10^{-0.23}\tau^{-0.1}_6      &  \tau\le 10^{5.2} \mbox{yr} \\
10^{-0.55}\tau^{-0.5}_6      &  \tau > 10^{5.2} \mbox{yr}\;,
\end{array} \right.
\label{Tc1}
\end{equation}
where $\tau_6$ is the age in units of $10^6$ yr. Based on the
cooling model derived from Tsuruta (1998) (in this model, the
effects of the magnetic field on thermal conductivity are
included, but polar cap heating is not considered), Hibschman \&
Arons (2001) use a temperature model as follow
\begin{equation}
T_{c,6}(t)=\left\{ \begin{array}{ll}
10^{0.05}\tau^{-0.1}_6      & \tau< 10^7 \mbox{yr} \\
10^{0.325}\tau^{-0.375}_6      & \tau> 10^7 \mbox{yr} \;,
\end{array} \right.
\label{Tc2}
\end{equation}
where $t$ is the spin-down age of the neutron star. For the hotter
component originating from the bombardment by high-energy
particles, the high-energy particles are produced either in the
polar cap region or in outer gap region, resulting in different
properties of the hotter component. In other words, the origin of
the relativistic particles, which bombard the stellar surface to
produce thermal hotter X-rays, depends on the models.

\subsubsection{X-ray emission from polar cap heating}
In the polar cap models, the return relativistic particles are
produced in the polar gap (e.g. Ruderman \& Sutherland 1975; Arons
1981). Recently, Zhang \& Harding (2000) considered full polar cap
cascade scenario of the polar cap model and estimated thermal

X-ray luminosity using a self-consistent polar cap heating in the
Harding \& Muslimov (1998) model (a relevant recent study see
Hibschman \& Arons 2001). Further, Harding \& Muslimov (2001)
studied the effect of pulsar polar cap heating produced by

positron returning from upper pair formation front, the polar cap
heating produces a thermal X-ray emission with a temperature,
$T_{pc}=(L_{pc} /\sigma A)^{1/4}$,
\begin{equation}
T_{pc,6}=\left\{ \begin{array}{ll}
2.46\left({P_{0.1}\over\tau_6}\right)^{{1\over28}}\left({\cos^2\alpha\over
a(\alpha)}
\right)^{{1\over4}}  & \mbox{for} P^{{9\over4}}_{0.1}<0.5B_{12} \\
2.51P^{{1\over8}}_{0.1}\left({\cos^2\alpha\over
a(\alpha)}\right)^{{1\over4}} & \mbox{for}
P^{{9\over4}}_{0.1}>0.5B_{12} \ \;,
\end{array} \right.
\label{Tpc}
\end{equation}
where $L_{pc}$ is the X-ray luminosity emitted from the polar
caps, $\sigma$ is the Stefan constant, and $A$ is the heated area
of the returning positrons from the polar gaps. For the canonical
polar cap area, $A=A_{pc}=\pi R^2(\Omega R/c)$. In the above
expressions, $P_{0.1}=P/0.1$ s, $B_{12}=B/10^{12}$ G and
$\tau_6=10^6\tau$ yr is the pulsar age, we have used the analytic
estimate of X-ray luminosity given by Harding and Muslimov (2001)
and $A_{pc}$, so the above expressions are valid for normal
pulsars with $\tau \le 10^7$ yr.

\subsubsection{X-ray emission from outer gap heating}

In the outer gap models, part of the relativistic particles from
the outer gap will collide the stellar surface, producing the
thermal X-rays. These relativistic inflowing particles from the
outer gap radiate away much of their energy before reaching the
polar cap (Zhang \& Cheng 1997, Zhu et al. 1997; Wang et al. 1998;
Cheng \& Zhang 1999). The residual energy of the charged particles
striking the polar cap is
\begin{equation}
E_e(R)\approx \left({2e^2c\over mc^3R_L}ln{r\over
R}\right)^{-1/3}\;\;.
\end{equation}
These particles collide with the polar cap at a rate of
$\dot{N}_e=f\dot{N}_{GJ}$, where $\dot{N}_{GJ}$ is the
Goldreich-Julian particle flux (Goldreich \& Julian 1969) and
estimated by Eq. (\ref{NGJ}).
 Therefore the polar cap is heated and radiates X-rays with a luminosity
($L_X\approx fE_e(R)\dot{N}_{GJ}$)
\begin{eqnarray}
L_X&\approx& 2.3\times 10^{31}fB_{12} P^{-5/3}\nonumber\\
& &R^3_6\left(\ln{r\over R}\right)^{-1/3}a(\alpha)\cos\alpha
\;\;{\mbox{ergs~s$^{-1}$}}\;\;.
\end{eqnarray}
These X-rays have a typical temperature
$T_h=(L_X/A\sigma_{SB})^{1/4}$ with
\begin{eqnarray}
T_h&\approx& 5.0\times 10^6P^{-1/6}B^{1/4}_{12} \left(\ln{r\over
R}\right)^{-1/12}\nonumber\\
& &(\sin^2(a(\alpha)\cos\alpha)^{1/4} \;\;{\mbox{K}}\;\;,
\label{Th}
\end{eqnarray}
where $A\sim \pi f\Omega R^3/c$ is the area of the stellar surface
bombarded by the return current
and $\sigma_{SB}$ is Stefan's constant. Generally, a part of these
X-rays will escape along the open magnetic field lines. Following
Cheng \& Zhang (1999), we use $L^h_X$ as the luminosity of the
X-rays escaping along the open field lines and introduce a
parameter $\xi=L^h_X/L_X$ (the estimate of $\xi$ see Cheng \&
Zhang (1999)). Most of X-rays are reflected back to the stellar
surface because of the cyclotron resonant X-ray reflecting mirror.
This process transfers emitted polar cap X-ray energy to the
entire surface of the neutron star. Finally, the X-rays from the
entire surface of the neutron star are emitted with a temperature
of $T=(L^s_X/2\pi R^2\sigma)^{1/4}$, where $L^s_X=(1-\xi)L_X$.
Corresponding characteristic temperature is $T_s=(L^s_X/ 4\pi
R^2\sigma_{SB})^{1/4}$, which gives
\begin{eqnarray}
T_s&\approx& 4.2\times 10^5(1-\xi)^{1/4}
f^{1/4}P^{-5/12}B^{1/4}_{12}\nonumber\\
& & \left(\ln{r\over R}\right)^{-1/12}(a(\alpha) \cos\alpha)^{1/4}
R^{1/4}_6 \;\;{\mbox{K}}\;\;. \label{Ts}
\end{eqnarray}

\subsubsection{Average energy of X-rays}

We consider two possible cases for the thermal X-rays from the
stellar surface. In the first case, the thermal X-ray are produced
by both the neutron star standard cooling mechanism and polar cap
heating, the temperatures are given by Eq. (\ref{Tc1}) or Eq.
(\ref{Tc2}) for the standard cooling mechanism and Eq. (\ref{Tpc})
for the polar cap heating. In the second case, the thermal X-ray
come from the bombardment of the relativistic particles from the
outer gap (e.g. Zhang \& Cheng 1997), the corresponding
temperatures are given by Eqs. (\ref{Ts}) and (\ref{Th}). Assuming
these X-ray can be approximated as the black-body, their spectrum
can be expressed as
\begin{equation}
F_X(E_X)=C\left[{E^2_X\over e^{E_X/kT_1}-1}+{A\over 4\pi R^2}
{E^2_X\over e^{E_X/kT_2}-1}\right]\;\;\;,
\end{equation}
where $R$ is the stellar radius, $A$ is the area of polar cap
being heated, and $T_1$ and $T_2$ are the temperatures being whole
stellar surface and polar cap respectively. Using above
expression, we can estimate the average X-ray energy as
\begin{eqnarray}
<E_X>&=&{\int F_X(E_X)E_XdE_X\over \int F_X(E_X)dE_X} ={\pi^4\over
30\zeta(3)}\nonumber\\ & &{1+(A/4\pi R^2)(T_2/T_1)^4\over
1+(A/4\pi R^2)(T_2/T_1)^3}(kT_1) \;\;\;. \label{Ex}
\end{eqnarray}
where $\zeta(x)$ is the Zeta function and $\zeta(3)\approx 1.2$.
For the different mechanisms of thermal X-rays from the stellar
surface, the dependence of $<E_X>$ on the basic parameters of
pulsars is different.

We consider two possible cases for estimating the average X-ray
energy: (i)thermal X-rays are produced by the standard neutron
star cooling and polar cap heating, we have
\begin{equation}
<E^{pc}_X>\approx 2.7kT_c{1+(R/4R_L)(T_{pc}/T_c)^4\over
1+(R/4R_L)(T_{pc}/T_c)^3} \;\;\;, \label{Expc}
\end{equation}
and (ii)the thermal X-rays are produced by the bombardment of the
relativistic particles from the outer gap, we have
\begin{equation}
<E^{og}_X>\approx 2.7kT_s{2-\xi\over1-\xi} \;\;\;. \label{Exog}
\end{equation}

\subsection{The Fractional Size of an Outer Gap}

In two dimensional geometry, the fractional size of the outer gap
is an important parameter for the $\gamma$-ray production in the
outer gap. According to Zhang \& Cheng (1997), the parallel
electric field in the outer gap can be approximated as
\begin{equation}
E_{||}=f^2B(r)\left({s\over R_L}\right)\;\;\;,
\label{Eparallel}
\end{equation}
where $f$ is the fractional size of the outer gap,
$B(r)=B_p(1+3\cos^2(\theta-\alpha))^{1/2}R^3/r^3$ is the
magnetic field strength at the radius, $r$, to the star, $R_L$ is the
radius of the light cylinder and $s$ is the curvature radius which is
(Lesch et al. 1998)
\begin{equation}
s(\theta,\theta_s)={R\over 3}{\sin(\theta-\alpha)\over
\sin^2(\theta_s- \alpha)}
{(1+3\cos^2(\theta-\alpha))^{{3\over2}}\over
1+\cos^2(\theta-\alpha)}\;\;\;, \label{cradius1}
\end{equation}
where $\theta_s$ is the polar angle at the stellar surface.
Equation (\ref{cradius}) can be written by
\begin{equation}
s(\theta,\theta_s)=\sqrt{rR_L}W(\alpha, r) \label{cradius}
\end{equation}
with
\begin{equation}
W(\alpha, r)={4\over 3}{[1-{3\over 4}a(\alpha){r\over
R}]^{3/2}\over \sqrt{a(\alpha)}(1-{1\over2}a(\alpha){r\over R_L})}
\end{equation}
where $a(\alpha)=\sin^2(\theta_c-\alpha)\sin\theta_c$. This
electric field will accelerate the electrons/positrons to
relativistic energy in the outer gap. Because these accelerated
particles will lose their energy through curvature radiation,
their Lorentz factor is estimated by using
$eE_{||}c=(2/3)e^2c\gamma^4/s^2$, which gives
\begin{eqnarray}
\gamma(r)&\approx& 2.84\times
10^7f^{{1\over2}}B^{{1\over 4}}_{12}P^{-{1\over 4}}R^{{3\over 4}}_6\left({r\over R_L}\right)^{-{3\over 8}}\nonumber\\
& &\left({\sqrt{1+3\cos^2(\theta-\alpha)}\over 2}\right)^{{1\over
4}} \;\;\;, \label{gamma}
\end{eqnarray}
where $P$ is the pulsar period in units of second and $R$ is the
stellar radius in units of $10^6$ cm. The characteristic energy of
the $\gamma$-ray photons in the outer gap can be approximated as
\begin{eqnarray}
E_{\gamma}&\approx& 143 f^{{3\over2}}B^{{3\over4}}_{12}P^{-{7\over
4}} \left({\sqrt{1+3\cos^2(\theta-\alpha)}\over 2}\right)^{{3\over
4}}
\nonumber\\
& &\left({r\over R_L}\right)^{-{3\over 2}} \left({s\over
R_L}\right)^{-{1\over4}}R^{{9\over4}}_6\;\;\; {\mbox{MeV}}\;\;\;,
\label{Ecur}
\end{eqnarray}

Inside the outer gap, the curvature photons interact with the thermal
X-rays from the stellar surface to produce $e^{\pm}$ pairs through
photon-photon pair production process, sustaining the outer gap. This pair
production condition is
\begin{equation}
<E_X>E_{\gamma}(1-\cos(\theta_{X\gamma}))=2(m_ec^2)^2\;\;\;,
\label{paircond}
\end{equation}
where $<E_X>$ is the average X-ray energy which is estimated
below, $\theta_{X\gamma}$ is the angle between the emission
directions of curvature photons and the thermal X-rays. We assume
that the curvature photons are emitted along the negative
direction of the magnetic field and the  thermal X-rays along the
radial direction, then we have
\begin{equation}
\cos\theta_{X\gamma}=
-{2\cos(\theta-\alpha)\over(3\cos^2(\theta-\alpha)+1)^{1/2}}\;\;,
\label{thetaXgamma}
\end{equation}
where $\theta$ is the polar angle at radius $r$. Putting Eq.
(\ref{sinlaw}) into Eq.(\ref{thetaXgamma}), we have
\begin{equation}
\cos\theta_{X\gamma}=-\left({1-(r/R_L)a(\alpha) \over
(1-(3/4)(r/R_L)a(\alpha))}\right)^{1/2}\;\;\;,
\end{equation}
where $a(\alpha)=\sin^2(\theta_c-\alpha)\sin\theta_c$.

We consider the fractional sizes of the outer gap corresponding to
two possible average energies of X-rays. In the first case, X-rays
are produced by the neutron star cooling and polar cap heating.
The average X-ray energy is given by Eq. (\ref{Expc}), and the
fractional size of the outer gap is
\begin{equation}
f(r,\alpha)\approx
6.9B^{-{1\over2}}_{12}P^{{7\over6}}<E_X>_{0.1}^{-{2\over3}}G_{pc}(r,\alpha)\;\;\;
\label{fsizepc}
\end{equation}
with
\begin{eqnarray}
G_{pc}&=&W^{1/6}\left({2\over 1-\cos\theta_{x\gamma}}\right)^{2/3}
\left({r\over R_L}\right)^{{13\over12}}\nonumber\\
& &\left({\sqrt{1+3\cos^2(\theta-\alpha)}\over
2}\right)^{-{1\over2}} \;\;, \label{Gpc}
\end{eqnarray}
where $<E_X>_{0.1}=<E_X>/0.1$ keV. Because the temperature of the
polar cap heating is greater than that of the neutron star
cooling, equation (\ref{Expc}) can be approximated as
$<E_X>\approx 2.7kT_c$. Using Eqs. (\ref{Tc1}) and (\ref{Tc2})
respectively, we have
\begin{equation}
<E_X>_{0.1}\approx\left\{
\begin{array}{ll}
32.9\left({P\over\dot{P}}\right)^{-0.1}     & {\scriptstyle\mbox{for} \dot{P}_{-15}\ge 10^{2}P }\\
5.2\times 10^6\left({P\over\dot{P}}\right)^{-0.5}
&{\scriptstyle\mbox{for} \dot{P}_{-15}< 10^{2}P} \;\;
\end{array} \right.
\label{EXpc1}
\end{equation}
and
\begin{equation}
<E_X>_{0.1}\approx\left\{
\begin{array}{ll}
7.49\left({P\over\dot{P}}\right)^{-0.11}   &{\scriptstyle \mbox{for} \dot{P}_{-15}\ge 1.6P} \\
7.33\times 10^4\left({P\over\dot{P}}\right)^{-{15\over40}}
&{\scriptstyle\mbox{for} \dot{P}_{-15}< 1.6P} \;\;,
\end{array} \right.
\label{EXpc2}
\end{equation}
where $\dot{P}=10^{-15}\dot{P}_{-15}$ is the period derivative of
a pulsar in units of s~s$^{-1}$. In the second case, the thermal
X-ray come from the bombardment of the relativistic particles from
the outer gap. Putting Eqs. (\ref{Ecur}) and (\ref{Ex}) into Eq.
(\ref{paircond}), we have
\begin{equation}
f(r,\alpha)\approx 5.2B^{-4/7}_{12}P^{26/21}R^{10/7}_6 G(r,
\alpha) \;\;\; \label{fsizeog}
\end{equation}
with
\begin{eqnarray}
G(r, \alpha)&=&\left[{2\over
1-\cos\theta_{X\gamma}}\right]^{{4\over7}} \left(\ln{r\over
R}\right)^{{1\over21}} \left({r\over
R_L}\right)^{{13\over14}}\nonumber\\
& &\left({W\over a(\alpha)\cos\alpha}\right)^{{1\over7}}
\left({\sqrt{1+3\cos^2(\theta-\alpha)}\over 2}\right)^{-{3\over7}}
\;\;\;.
\end{eqnarray}
Obviously $f$ is a function of $r$ as well as the inclination
angle $\alpha$ in the two cases.

It is believed that an outer gap start at the null charge surface
(${\bf{\Omega}}\cdot {\bf{B}}=0$), which defines the inner
boundary of the outer gap and the radial distance is $r_{in}$.
From Eq. (\ref{fsizepc}) or Eq. (\ref{fsizeog}), the fractional
size reaches a minimum at the radius ($r_{in}$) of the inner
boundary, and then increases with radius for a given pulsar.
Therefore, the fractional size of the outer gap at the radius
$r_{in}$ determine whether or not the outer gap exists. If
$f(r_{in},\alpha)>1$, it means that the pulsar does not exist any
outer gap. In other words, a pulsar with $f(r_{in},\alpha)\le 1$
should have its outer gap and will emit high-energy photons
produced in the outer gap. As the radius increases, the fractional
size of the outer gap increases. For a pulsar with
$f(r_c,\alpha)\le 1$, the outer gap will extend from $r_{in}$ to
$r_c$. However, the pulsar with $f(r_c,\alpha)> 1$ will stop at
some radius $r_b$, in which $f(r_b,\alpha)=1$.  In order to
explain the average properties of high-energy photon emission from
the outer gap, we assume that high-energy emission at a average
radius $<r>$ represents the typical emission of high-energy
photons from a pulsar. The average radius is given by
\begin{equation}
<r>={\int^{r_{max}}_{r_{in}}f(r,\alpha)rdr \over
\int^{r_{max}}_{r_{in}}f(r,\alpha)dr}\;\;, \label{rave}
\end{equation}
where $r_{max}=min(r_c,r_b)$.  The average gap size is
approximated as $f(<r>, P, B)$ by substituting $<r>$ into Eq. (32)
and Eq. (36) and in general a function of $P$, $B$ and $\alpha$
because $r_c$ is a function of $P$ and $\alpha$, and $r_b$ are
function of $P$, $B$ and $\alpha$.

Generally, the inclination angles for the pulsars are not known
well, so we consider the changes of the fractional size with the
inclination angle. Let us define the following function:
\begin{equation}
f(<r>, P, B) = \eta (\alpha, P, B) f_o (P, B)
\end{equation}
where $f_o (P, B) = 5.5 P^{26/21} B^{-4/7}_{12}$ is the fractional
size of outer gap by ignoring the effect of inclination angle
(Zhang \& Cheng 1997). So the effect of inclination angle should
exhibit in the function $\eta (\alpha, P, B)$. In Fig. 1, we
consider the variation of outer gap size with magnetic inclination
angle for various $P$ and $B$. In panel A of Fig. 1, the upper
(lower) solid line is the self-consistent outer gap model (the
cooling X-ray outer gap model) with $P = 0.1s$ and the upper
(lower) dashed line is the self-consistent outer gap model (the
cooling X-ray outer gap model) with $P = 0.3s$ respectively. The
magnetic field is chosen to be $3\times 10^{12}$G. In panel B of
Fig. 1, the upper (lower) solid line is the self-consistent outer
gap model (the cooling X-ray outer gap model) with $B_{12} = 3.0$
and the upper (lower) dashed line is the self-consistent outer gap
model (the cooling X-ray outer gap model) with $B_{12} = 1.5$
respectively. The period is chosen to be 0.2s. We can see that
$\eta$ varies about a factor of 2 for the self-consistent model
but becomes rather constant for the cooling X-ray outer gap model.

\section{High-Energy Gamma-ray Luminosity}

EGRET has observed six pulsars with high confidence and three
possible radio pulsars to emit high-energy gamma-rays above 100
MeV (see Thompson 2001 for a review). One of these possible radio
pulsars is a millisecond pulsar PSR J0218+4232 (Kuiper et al.
2000). Therefore, there could be eight young pulsars to emit
high-energy gamma-rays observed by EGRET. It should be noted that
PSR B1509-58 is also a gamma-ray pulsar. But it is seen only up to
10 MeV by COMPTEL (Kuiper et al. 1999) and not above 100 MeV by
EGRET. The observed gamma-ray luminosity of a pulsar is
$L^{obs}_{\gamma}=4\pi d^2\zeta F_{\gamma}$, where $F_{\gamma}$ is
the observed gamma-ray flux, $d$ is the distance to the pulsar and
$\zeta$ is the gamma-ray beaming fraction ($0<\zeta\leq 1$), which
is defined as the ratio of the beaming solid angle to $4\pi$. Two
parameters are highly uncertain: the distance $d$ and the beaming
fraction $\zeta$. It is commonly assumed that $\zeta=1/4\pi$ in
estimating the observed gamma-ray luminosity. However, the
$\gamma$-ray beaming fraction should be different for various
$\gamma$-ray pulsars, which is a function of the magnetic
inclination angle as well as the size of the outer gap. Some
approximation forms of beaming fraction have been
given(Yadigaroglu \& Romani 1995; Romani 1996; Zhang, Zhang \&
Cheng 2000). How accurate of these approximation forms are not
known. Furthermore, we want to emphasize that in addition the
beaming fraction the distance estimate also affects the observed
gamma-ray luminosity strongly. For example, the recent measurement
shows that the distance to the Vela pulsar is $294^{+76}_{-50}$ pc
(Cavareo et al. 2001), which is less than previous value (500 pc)
derived from radio observation. Hence the gamma-ray luminosity of
the Vela could be a factor of 2 lower than that given in Thompson
et al.(2001). In this paper, we just want to see how the
inclination affects  the gamma-ray luminosity. For simplification,
we use a common assumption of $\zeta=1/4\pi$ in order to compare
with the observed data given by Thompson et al.(2001).

Because high-energy gamma-rays are mainly produced from the outer
gap in our model, we will compare our expected gamma-ray
luminosities with those of the gamma-ray pulsars which emit
gamma-rays observed by EGRET. In our model, gamma-ray luminosity
for each gamma-ray pulsar depends on period, magnetic field and
the magnetic inclination. However, the magnetic inclination angles
are not known well. Once the average fractional size of the outer
gap for a pulsar is estimated, the gamma-ray luminosity can be
approximated as
\begin{equation}
L_{\gamma}\approx f^3(<r>)L_{sd}\;\;\;, \label{Lgamma}
\end{equation}
where $L_{sd}=4\pi^2I\dot{P}/P^{3}$ is the spin-down luminosity of
the pulsar and $I=10^{45}$ g~cm$^2$. Using Eq. (\ref{Bp}) and
putting Eq. (\ref{fsizeog}) into Eq. (\ref{Lgamma}), we have
\begin{equation}
L_{\gamma}= L_{\gamma, 0}\eta^3(\alpha, P, B) \label{Lgamma1}
\end{equation}
with
\begin{equation}
L_{\gamma, 0}\approx 1.36\times 10^{33}
B^{2/7}_{12}P^{-2/7}\;\;\;. \label{Lgamma0}
\end{equation}
It is obviously that Eq. (\ref{Lgamma0}) is the same as that given
by Zhang \& Cheng (1997).

 In principle, we can estimate $\gamma$-ray
luminosity for a pulsar with a known inclination angle. However,
it is difficult to estimate $\gamma$-ray luminosity for all
canonical pulsars because only the inclination angles of a few of
pulsars are known. Therefore, we find out statistically the
relation between $\gamma$-ray luminosity and pulsar spin-down
power for the canonical pulsars using Monte Carlo method. The
details of this Monte Carlo method is given by Cheng \& Zhang
(1998) and Zhang, Zhang \& Cheng (2000). We use the following
assumptions for generating the Galactic pulsar population:
\begin{enumerate}
\item [(i)] The pulsars are born at a rate ($\dot{N}_{NS}\sim
(1-2)$ per century) with spin periods of $P_0=10$ ms.

\item [(ii)] The initial
position for each pulsar is estimated from the distributions
$\rho_z(z)=(1/z_{\rm exp})\rm exp(-|z|/z_{exp})$ and
$\rho_R(R)=(a_R/R^2_{\rm exp})$ $R~\rm exp(-R/R_{exp})$,
 where $z$ is the distance from the Galactic plane, $R$ is the distance
from the Galactic center, $z_{exp}=75$ pc, $a_R=[1-e^{-R_{\rm
max}/R_{\rm exp}}(1+R_{\rm max}/R_{\rm exp})]^{-1}$,

 $R_{\rm exp}=4.5$ kpc and $R_{\rm max}=20$ kpc (Paczynski 1990; Sturner
\& Dermer 1996).

\item [(iii)] The initial magnetic fields are distributed as
a Gaussian in $\log$B with a mean $\log B=12.52$ and a dispersion
$\sigma_B=0.35$. We ignore any field decay for these
rotation-powered pulsars.

\item [(iv)] The
initial velocity of each pulsar is the vector sum of the circular
rotation velocity at the birth location and a random velocity from
the supernova explosion (Paczynski 1990; Cheng \& Zhang 1998), the
circular velocity is determined by Galactic gravitational
potential and the random velocity are  distributed as a Maxwellian
distribution with dispersion of three dimensional velocity
$\sigma_V=\sqrt{3}\times 100$ km/s (Lorimer, Bailes \& Harrison
1997).

\item [(v)] A random distribution of magnetic inclination angles
is used (e.g. Biggs 1990). However, the values of $\alpha$ subject
to two constraints. First the photon energy given in Eq. (28)
cannot higher than the total potential drop of the outer gap.
Second $r_{in}$ in Eq. (9) must be larger than the stellar radius.
\end{enumerate}

The pulsar period at time $t$ can be estimated by
\begin{equation}
P(t)=\left[P^2_0+\left({16\pi^2 R^6_{NS}B^2\over 3I
c^3}\right)t\right]^{1/2}\;\;, \label{period}
\end{equation}
where, $R_{NS}$ is the neutron star radius and $I$ is the neutron
star moment of inertia. The period derivative ($\dot{P}$) can be
determined by
\begin{equation}
P\dot{P}=(8\pi^2R^6_{NS}/3Ic^3)B^2\;\;\;. \label{pdot}
\end{equation}
Furthermore, the pulsar position at time $t$ is determined
following its motion in the Galactic gravitational potential.
Using the equations given by Paczynski (1990) for given initial
velocity, the orbit integrations are performed by using the 4th
order Runge Kutta method with variable time step \cite{press92} on
the variables $R$, $V_R$, $z$, $V_Z$ and $\phi$. Then the sky
position and the distance of the simulated pulsar can be
calculated.

We now consider the observational selection effects. First, we
consider radio selection effects in order to generate a pulsar
population detectable at the radio band: the pulsar must satisfy
that its radio flux is greater than the radio survey flux
threshold and its broadened pulse width is less than the rotation
period (e.g. Sturner \& Dermer 1996). The pulsar which satisfies
\begin{equation}
L_{400}/d^2\ge S_{\rm min} \label{Smin}
\end{equation}
is considered to be a radio-detectable pulsar, where $L_{400}$ is
the radio luminosity at 400 MHz and $d$ is the distance to the
pulsar. The radio beaming fraction can be expressed as (Emmering
\& Chevalier 1989)
\begin{equation}
f_r(\omega)=(1-\cos\omega)+(\pi/2-\omega)\sin\omega\;\;,
\label{f_radio}
\end{equation}
where $\omega=6^{\circ}.2\times P^{-1/2}$ (e.g. Biggs, 1990) is
the half-angle of the radio emission cone. Then, following
Emmering \& Chevalier (1989), a sample pulsar with a given period
$P$ is chosen in one out of $f_r(P)^{-1}$  cases using the Monte
Carlo method. Second, we consider the $\gamma$-ray selection
effects. According to Zhang, Zhang and Cheng (2000), we use
\begin{equation}
S_{\gamma}(>100 {\hbox{MeV}}) \ge 1.2\times
10^{-10}\;\;{\hbox{erg~cm$^{-2}$s$^{-1}$}} \label{gammacond}
\end{equation}
as the minimum detectable $\gamma$-ray energy flux, which
corresponds roughly to the faintest sources with $(TS)^{1/2}>5$.

Using above method, we can generate a $\gamma$-ray pulsar
population. In Fig. 2, we show the relation between $\gamma$-ray
luminosity and the spin-down power for the simulated $\gamma$-ray
pulsar population by using this new outer gap model, where the
birth rate of the neutron stars is assumed to be 1/200 yr and the
pulsar ages are limited to be not greater than $10^7$ yr. In this
figure, we show two pulsar populations, one is the population
which radio selection effects of the pulsar with outer gaps are
taken into account (i.e. radio-loud $\gamma$-ray pulsars see
shaded circles in Fig.2), and the other population is the
radio-quiet $\gamma$-ray pulsars (see open circles in Fig.2) .

There are several interesting features in this figure. First there
is a rather sharp boundary on the left of the population. In fact
this boundary is given by
\begin{equation}
L_{\gamma} = L_{sd}.
\end{equation}
This relation/boundary results from the fact that $\gamma$-ray
pulsars terminate at $f(<r>)=1$ and $r_{max} = r_b$ in Eq. (38).
Similarly the second rough boundary appearing at the bottom of the
population is caused by $f(<r>)=1$ and $r_{max} = r_c = r_L$.
Zhang \& Cheng (1997) has estimated the fractional gap size by
assuming that the typical distance from the gap to the star is
$\sim r_L$, they obtained $f = 5.5 B_{12}^{-4/7} P^{26/21}$. When
we substitute this relation into $L_{sd}$, we obtain
\begin{equation}
L^{crit}_{sd} = 1.5 \times 10^{34} P^{1/3} erg s^{-1}.
\end{equation}
It is important to note that $L_{\gamma} = L_{sd}$ as $L_{sd}
\lesssim L^{crit}_{sd}$.

In such a pulsar population, the best fit gives following relation
\begin{equation}
\log L_{\gamma}\approx 20.42+0.38\log L_{sd} \label{LgEGRET}
\end{equation}
for the pulsars with outer gaps. When taking the radio selection
effects, the slope between $L_{\gamma}$ and $L_{sd}$ becomes
flatter, which is $\log L_{\gamma}\approx 23.25+0.30\log L_{sd}$.
We will consider the statistics of the pulsars with outer gaps in
the rest of this section. In order to show the important effect of
the magnetic inclination angle on $\gamma$-ray luminosity, we show
the change of $\log_{10}(L_{\gamma}/L_{\gamma, 0})$ with
$\log_{10} L_{\gamma}$ in Fig. 3. The best fit indicates that
$\log_{10}(L_{\gamma}/L_{\gamma, 0})\propto
0.9\log_{10}L_{\gamma}$. For comparison, we show the result given
by Zhang \& Cheng (1997), which is independent with the
inclination angle.

It is very interesting that the slope between $L_{\gamma}$ and
$L_{sd}$ will become steeper when the minimum detectable
$\gamma$-ray energy flux decreases. For example, we obtain a
$\gamma$-ray pulsar population using GLAST threshold ($1.8\times
10^{-12}$ erg~cm$^{-2}$~s$^{-1}$), the best fit gives
\begin{equation}
\log L_{\gamma}\approx 17.45+0.46\log L_{sd}\;\;\;.
\label{LgGLAST}
\end{equation}
That is $L_{\gamma}\propto L^{1/2}_{sd}$. In this pulsar
population, we have $\log_{10}(L_{\gamma}/L_{\gamma, 0})\propto
0.92\log L_{\gamma}$.

For comparison, we show our results given by Eqs. (\ref{LgEGRET})
and (\ref{LgGLAST}) and the observed high-energy photon
luminosities of eight $\gamma$-ray pulsars in Fig. 4. In this
figure, the observed data are taken from Thompson (2001), which
derived from detected fluxes above 1 eV assuming the solid angle
of photon beaming of 1 sr. From Fig. 5, it can be seen that both
our results given by Eqs. (\ref{LgEGRET}) and (\ref{LgGLAST}) are
consistent with the observed data. For the result fitting the
simulated $\gamma$-ray pulsar population with EGRET threshold, the
expected slope ($0.38$) is flatter than the observed one ($0.46$).

Finally we would like to point out that this new outer gap model
predicts many more weak $\gamma$-ray pulsars whose $\gamma$-ray
luminosities can be as low as $10^{32} erg ~s^{-1}$ (cf. Fig. 2)
than previous outer gap models (Zhang \& Cheng 1997; Cheng \&
Zhang 1998). The main reason for existing these weak $\gamma$-ray
pulsars is that instead of using the outer gap size at half of
light cylinder radius to determine if the outer gap can exist, the
new model allows the outer gap to survive when the fractional size
of outer gap at the null charge surface for a given pulsar is less
than unity. This prediction allows the age of $\gamma$-ray pulsars
to extend to a few million years old. In fact, more detail Monte
Carlo simulation results show that most of these weak $\gamma$-ray
pulsars have ages between 0.3-3 million years old and many of them
are located at higher galactic latitude. Most of these weak
$\gamma$-ray pulsars their radio beam could miss the Earth and
they could contribute to the unidentified $\gamma$-ray sources in
high galactic latitude (Cheng et al. 2003). These predictions can
be verified by GLAST.

\section{The Death lines of Pulsars with Outer Gaps}

We now consider the condition which the outer gap of a pulsar
exists. For an outer gap, the inner boundary is estimated as the
null charge surface where the magnetic field lines are
perpendicular to the rotation axis. From Eq. (\ref{fsizeog}) or
(\ref{fsizepc}), $f(r, \alpha)$ reaches minimum at $r=r_{in}$. In
other words, if the fractional size of the outer gap at $r_{in}$
is larger than unity, then the outer gap would not exist.
Therefore, we can estimate the death lines of the pulsars with
outer gaps by using $f(r_{in}, \alpha)=1$. It should be noted that
$G_{pc}(r_{in}, \alpha)$ in Eq. (\ref{fsizepc}) or $G(r_{in},
\alpha)$ in Eq. (\ref{fsizeog}) is only the function of $\alpha$
because $r_{in}/R_L$ only depends on $\alpha$ (see Eq.
(\ref{rin})). For the case of X-rays produced by the neutron star
cooling and polar cap heating, we obtain from $f(r_{in},
\alpha)=1$
\begin{equation}
\log \dot{P}=\left\{ \begin{array}{ll}
3.1\log P-A_1(\alpha)  &{\scriptstyle \mbox{for} \log P\le 13+\log \dot{P}} \\
{15\over 7}\log P-A_2(\alpha) &{\scriptstyle \mbox{for} \log P>
13+\log \dot{P}} \ \;
\end{array} \right.
\label{deathpc1}
\end{equation}

for the temperature given by Zhang \& Harding (2000), where
$A_1(\alpha)= 12.87-3.16\log G_{pc}(\alpha)$ and
$A_2(\alpha)=12.92-{12\over 7}\log G_{pc}(\alpha)$;  and
\begin{equation}
\log \dot{P}=\left\{ \begin{array}{ll}
3.1\log P-B_1(\alpha)  & {\scriptstyle\mbox{for} \log P\le 14.8+\log \dot{P}} \\

{7\over 3}\log P-B_2(\alpha)  & {\scriptstyle\mbox{for} \log
P>14.8+\log \dot{P}} \ \;
\end{array} \right.
\label{deathpc2}
\end{equation}
for the temperature used by Hibschman \& Arons (2001), where
$B_1(\alpha)= 13.46-3.16\log G_{pc}(\alpha)$ and

$B_2(\alpha)=13.95-2\log G_{pc}(\alpha)$. For the case of X-rays
produced by the bombardment of the returning particles from the
outer gap, we have
\begin{equation}
\log \dot{P}={10\over 3}\log P-13.02+{7\over 2}\log G(\alpha)\;\;.
\label{deathog}
\end{equation}
Obviously, the death lines depend on the magnetic inclination
angle.

Although the magnetic inclination angle of each pulsar has not
been determined well, the distribution of the magnetic inclination
angle can be estimated from the statistical polarization study of
the radio pulsars. It was generally believed that the parent
distribution of the magnetic inclinations satisfy an uniform
distribution (Gunn \& Ostriker 1970; Gil \& Han 1996). However,
recent study by using polarization data of the radio pulsars
indicate that the parent distribution of the magnetic inclinations
satisfies a cosine-like distribution (Tauris \& Manchester 1998).
Therefore, we estimate the average value of $f(r_{in}, \alpha)$ in
these two possible parent distributions of the magnetic
inclinations, i.e.
\begin{equation}
<G(\alpha)>=\int G(\alpha)U(\alpha)d\alpha/\int
U(\alpha)d\alpha\;\;,
\end{equation}
where $U(\alpha)$ represents the distribution of the inclination
angles. We consider uniform and cosine-like distributions of the
inclination angles respectively. For the uniform distribution, we
have $<G_{pc}(\alpha)>\approx 0.32$ and $G(\alpha)=0.38$.
Therefore Eqs. (\ref{deathpc1}), (\ref{deathpc2}) and
(\ref{deathog}) become
\begin{equation}
\log \dot{P}=\left\{ \begin{array}{ll}
3.1\log P-14.43 & {\scriptstyle\mbox{for} \log P\le 13+\log \dot{P}} \\
{15\over 7}\log P-13.77 &{\scriptstyle \mbox{for} \log P > 13+\log
\dot{P}} \;,
\end{array} \right.
\label{deathpcu1}
\end{equation}
\begin{equation}
\log \dot{P}=\left\{ \begin{array}{ll}
3.1\log P-15.02  & {\scriptstyle\mbox{for} \log P\le 14.8+\log \dot{P}} \\
{7\over 3}\log P-14.94  &{\scriptstyle \mbox{for} \log P>14.8+\log
\dot{P}} \;
\end{array} \right.
\label{deathpcu2}
\end{equation}
and
\begin{equation}
\log \dot{P}={10\over 3}\log P-14.60\;\;. \label{deathogu}
\end{equation}
For cosine distribution, $<G_{pc}(\alpha)>\approx 0.43$ and
$G(\alpha)\approx 0.49$, Eqs. (\ref{deathpc1}),(\ref{deathpc2})
and (\ref{deathog}) become
\begin{equation}
\log \dot{P}=\left\{ \begin{array}{ll}
3.1\log P-14.03 &{\scriptstyle \mbox{for} \log P\le 13+\log \dot{P}} \\
{15\over 7}\log P-13.55 &{\scriptstyle \mbox{for} \log P > 13+\log
\dot{P}} \ \;,
\end{array} \right.
\label{deathpccos1}
\end{equation}
\begin{equation}
\log \dot{P}=\left\{ \begin{array}{ll}
3.1\log P-14.62  &{\scriptstyle \mbox{for} \log P\le 14.8+\log \dot{P}} \\
{7\over 3}\log P-14.68  &{\scriptstyle \mbox{for} \log
P>14.80+\log \dot{P}} \;
\end{array} \right.
\label{deathpccos2}
\end{equation}
and
\begin{equation}
\log \dot{P}={10\over 3}\log P-14.20\;\;. \label{deathogcos}
\end{equation}

Based on the original outer gap model (CHR I; CHR II), Chen \&
Ruderman (1993) have considered the death lines of Crab-like and
Vela-like pulsars respectively. For the Crab-like pulsars, the
death line is (see Eq. (25) of their paper)
\begin{equation}
\log \dot{P}=4\log P- 7\;\;. \label{death3}
\end{equation}
For Vela-like pulsars, they introduced a parameter $\xi$ to
describe the expected variation of the magnetic field within the outer gap
and assumed $\xi=1/2$. In this case, the death line is (see Eq. (27) of
their paper)
\begin{equation}
\log \dot{P}=3.8\log P- 10.2\;\;.
\label{death4}
\end{equation}
According to Zhang \& Cheng (1997), we have from $f_s\approx
2.83\times 10^{-4}P^{20/21}\dot{P}^{-2/7}=1$
\begin{equation}
\log \dot{P}={10\over 3}\log P- 12.42\;\;. \label{death5}
\end{equation}

In Fig. 5, we show the death lines of the pulsars with outer gaps
for two possible distributions of the magnetic inclination angles.
In this case, we assume that X-rays are produced by the neutron
star cooling and polar cap heating. Because the temperature used
by Hibschman \& Arons (2001) is higher than that used by Zhang \&
Harding (2000), the average energy of X-rays for the former is
greater than that for the latter. Therefore, the death lines
obtained by using the temperature of Hibschman \& Arons (2001) are
lower than those by using the temperature of Zhang \& Harding
(2000).

In Fig. 6, we show the death lines of the pulsars with outer gaps,
in which X-rays are produced by the returning relativistic
particles from the outer gap, and the outer gap is self-sustained.
For comparison, we also plot the death lines given by Chen \&
Ruderman (1993), and the death line derived from the model of
Zhang \& Cheng (1997). It can be seen that our model predicts that
much more radio pulsars have self-sustained outer gaps compared to
those given by Chen \& Ruderman (1993) as well as by Zhang \&
Cheng (1997). In our estimate of the death lines of the pulsars
with self-sustained outer gaps, we require that $f(r_{in})=1$.
This condition is reasonable. It is believed that an outer gap can
develop along the null charge surface (Cheng, Ruderman \&
Sutherland 1976) or along the last closed field line (CHR I; CHR
II). Therefore, a self-sustained outer gap exists if the
fractional size of the outer gap at $r_{in}$ is not greater than
unity.

\section{Conclusion and Discussion}

After taking the geometry of the dipole magnetic field, we have
given a revised version of the outer gap model given by Zhang \&
Cheng (1997). In the revised outer gap model, the fractional size
of the outer gap is not only the function of period and magnetic
field of the neutron star, but also the function of the radial
distance ($r$) to the neutron star and the magnetic inclination
angle ($\alpha$). In other words, the fractional size of the outer
gap has a form of $f(r,\alpha)=f_i(P, B_{12})G_i(r,\alpha)$, where
$f_0(P, B_{12})$ is only the function of pulsar period and surface
magnetic field, $G_i(r,\alpha)$ changes with radial distance to
the neutron star and the magnetic inclination angle and subscript
$i$ represents the X-ray field considered. The fractional size of
the outer gap is given by Eq. (\ref{fsizepc}) in the X-ray field
which is produced by the neutron star cooling and polar cap
heating and by Eq. (\ref{fsizeog}) in the X-ray field which is
produced by outer gap heating. In this model, the fractional size
of the outer gap has a minimum at the inner boundary for a given
pulsar, increases with the radial distance along the last open
field lines and then reaches its outer boundary where $f(r_b,
\alpha)=1$. In other words, the outer gaps of some pulsars do not
start from the null charge surface to the light cylinder. We have
shown the changes of the fractional size of the outer gap with the

inclination angle (see Fig.1). Further, we have found that the
outer gaps of relative young pulsars such Vela can extend from the
null charge surface to the light cylinder for any inclination
angle , however, the outer gaps of some pulsars like as Geminga
cannot extend to the light cylinder for a larger inclination angle
(say $75^{\circ}$).

In order to describe the average properties of high-energy
radiation from a $\gamma$-ray pulsar, we have defined an average
radial distance $<r>$ (see Eq. (\ref{rave})). Using Monte Carlo
method described by Cheng \& Zhang (1998) (also see Zhang, Zhang
\& Cheng 2000), we simulated two populations of $\gamma$-ray
pulsars whose energy fluxes are greater than EGRET threshold and
GLAST threshold. In these simulations, we only consider the case
of pulsars with self-sustained outer gap and uniform distribution
of the inclination angles. We have plotted the change of
$L_{\gamma}$ with $L_{sd}$ (see Fig. 2). We also indicated the
variation of $L_{\gamma}/L_{\gamma,0}$ with $L_{\gamma}$ in Fig.
3, which show the importance of the inclination angle on
$L_{\gamma}$. In the model of Zhang \& Cheng (1997), $L_{\gamma}$
is independent on the inclination angle. Fitting the simulated
results, we found that $L_{\gamma}\propto L^{0.38}_{sd}$ for EGRET
threshold and $L_{\gamma}\propto L^{0.46}_{sd}$ for GLAST
threshold. Compared with the observed data given by Thompson
(2001), our simulated results are reasonable (see Fig. 4). In
fact, current distance of the Vela pulsar (Cavareo,  et al. 2001)
is less than that used by Thompson (2001) about a factor of two,
which reduce the luminosity about a factor of four.

In Fig.4, we note that Geminga is not within the simulated
population. Zhang \& Cheng (2001) have used a three-dimension
outer gap model to explain the phase-resolved spectra of Geminga.
They discovered that in order to fit the observed $\gamma$-ray
data the solid angle $\Delta \Omega$ must be near 5$sr$. In Fig.
4, $\Delta \Omega = 1sr$ is used for all observed $\gamma$-ray
pulsars. If the larger solid angle is used, it brings Geminga
within the simulated population. It should be pointed out that
Yadigaroglu \& Romani (1995) have studied the effect of
$\gamma$-ray beaming in their outer gap model (also see Romani
1996). Zhang et al (2000) also gave a approximate expression of
the $\gamma$-ray beaming fraction, which will be applied to
estimate $\gamma$-ray fluxes in our new model.

It is interesting to point out that both polar gap models and
outer gap models predict $L_{\gamma}=L_{sd}$ for low spin-down
power pulsars. But the position of the break occurs at $5 \times
10^{33} P^{-1/2} erg/s$ for the polar gap model (Harding et al.
2002) whereas the outer gap models predict a higher position at
$1.5 \times 10^{34} P^{1/3} ergs^{-1}$. For higher spin-down power
pulsars, the polar gap models predict $L_{\gamma}=L^{1/2}_{sd}$,
whereas the outer gap cannot give precise prediction because the
model $L_{\gamma}$ also depends on a less well-known parameter
$\alpha$. The statistical predictions of the outer gap model give
$L_{\gamma}=L^{\delta}_{sd}$, where $\delta$ depends on the
properties of $\gamma$-ray detector. For example, $\delta$ = 0.38
for the EGRET and $\delta$ = 0.46 for GLAST.

According to our model, the fractional size of outer gap at the
null charge surface for a given pulsar ($f(r_{in}, \alpha)$)
reaches a minimum. Therefore, the outer gap should exist only if
$f(r_{in},\alpha)\le 1$. Averaging $f(r_{in}, \alpha)$ on two
possible (uniform and cosine) distributions of the magnetic
inclination angles respectively, we have obtained the death lines
of the pulsars with outer gaps in the two possible X-ray fields
and the comparison them with the observed data in Figs.5 and 6.

Compared with the death line derived from the outer gap model of
Zhang \& Cheng (1997), the revised model predict that more pulsars
will have their outer gaps and then emit high-energy photons.

We would like to make the conclusion as follows. Our
results indicate that (i) the intrinsic parameters for explaining
the observed $\gamma$-ray properties of rotation-powered pulsars
are the magnetic inclination angle, period and magnetic field. We
have obtained a very concrete functional form of the prediction of
gamma-ray luminosity which only depends on  these intrinsic
parameters, the inclination angle could be known if the radio data
is sufficiently good; (ii) the conversion efficiency for 7 known
gamma-ray pulsars, that is rather scattered , can be explained in
our revised model using the Monte Carlo method. Although our
estimation of the outer gap size cannot precisely predict the
thickness of individual pulsar when the magnetic inclination angle
is poorly known, it is a reasonable estimation of the statistical
properties of gamma-ray pulsars using the statistical method;
(iii)Unlike those 7 observed gamma-ray pulsars, mature pulsars
(ages~0.3-3 million years) can also be gamma-ray pulsars and their
efficiency is insensitive to the inclination angle. Most
importantly, their gamma-ray luminosity is proportional to the
spin-down power, which can be tested by GLAST; (iv) The mean
cut-off age of gamma-ray pulsars is increased by a factor of 3,
therefore older gamma-ray pulsars (up to about 3 million years
old) can move up to higher galactic latitude. Some unidentified
EGRET gamma-ray sources could be mature pulsars with ages between
0.3-3 million years old pulsars predicted by this model. In fact,
Parkes Observatory survey has discovered a large number of radio
pulsars on the error boxes of EGRET unidentified gamma-ray point
sources (Torres et al. 2003); and (v)  The previous works on death
line based on the outer gap model (e.g. Chen and Ruderman 1993)
did not give a detail physical reason rather they used a
phenomenological approach. Here we proposed a new criterion based
on concrete physical reason for the death line, which predicts
more gamma-ray pulsars. (vi)Our model predicts many more weak
$\gamma$-ray pulsars with age between 0.3 – 3 million years old
than previous outer gap models and they satisfy $L_{\gamma}
\propto L_{sd}$, which can be verified by GLAST.

Finally, we would like to remark that the gamma-ray luminosity
formulae developed in this paper may not be able to explain the
gamma-ray luminosity of individual pulsar. Two important factors,
i.e. distance uncertainty and beaming fraction, which play crucial
roles in determining the gamma-ray luminosity, have not been
considered here. For example, the luminosity ratio between PSR
B1055-52 and Geminga, which have same spin-down power of $3\times
10^{34} erg~s^{-1}$, is about a factor of 6 (Thompson et al.
2001). But the model prediction (cf. Fig. 2) is no more than a
factor of 3. In fact, the estimate of the distance of PSR B1055-52
is very uncertain, its distance can change from $\sim$ 1.5 kpc
(for example, Thompson et al. 2001) to a small value of $\sim$500
pc (see, for example, \"{O}gelman \& Finley 1993; Combi et al.
1997; Torres, Butt, Camilo 2001; Hirotani \& Shibata 2002), which
makes the ratio change from  $\sim$ 10 to $\sim$ 3. Recent
distance estimate for PSR B1055-52 from the dispersion measure is
$\sim$ 0.72 kpc (see http:
//rsd-www.nrl.navy.mil/7213/lazio/ne\_model, also see
Mignani, DeLuca \& Caraveo 2003). However, it is well known that
25\% error is common in determining the dispersion measure
(MaLaughlin \& Cordes 2000), which gives a factor of 2
uncertainties in the luminosity. Also the distance estimate of
Geminga is known to have at least 25\% uncertainties (Caraveo et
al. 1996), which give another factor of 2 uncertainties in
gamma-ray luminosity estimate. Other uncertainty results from the
gamma-ray beaming fraction. In principle, both distance estimate
and beaming fraction will be obtained more accurate. Then our
model still predicts that the difference in the magnetic
inclination angle can still cause a large difference in gamma-ray
luminosity for pulsars with same spin-down power.

\acknowledgments We thank the anonymous referee for his/her very
constructive comments. This work is partially supported by
'Hundred Talents Program of CAS', National 973 projection of China
 (NKBRSFC 19990754) and the RGC grant of Hong Kong Government.

\clearpage

\begin{figure}
\vbox to2.1in{\rule{0pt}{2.1in}} \includegraphics{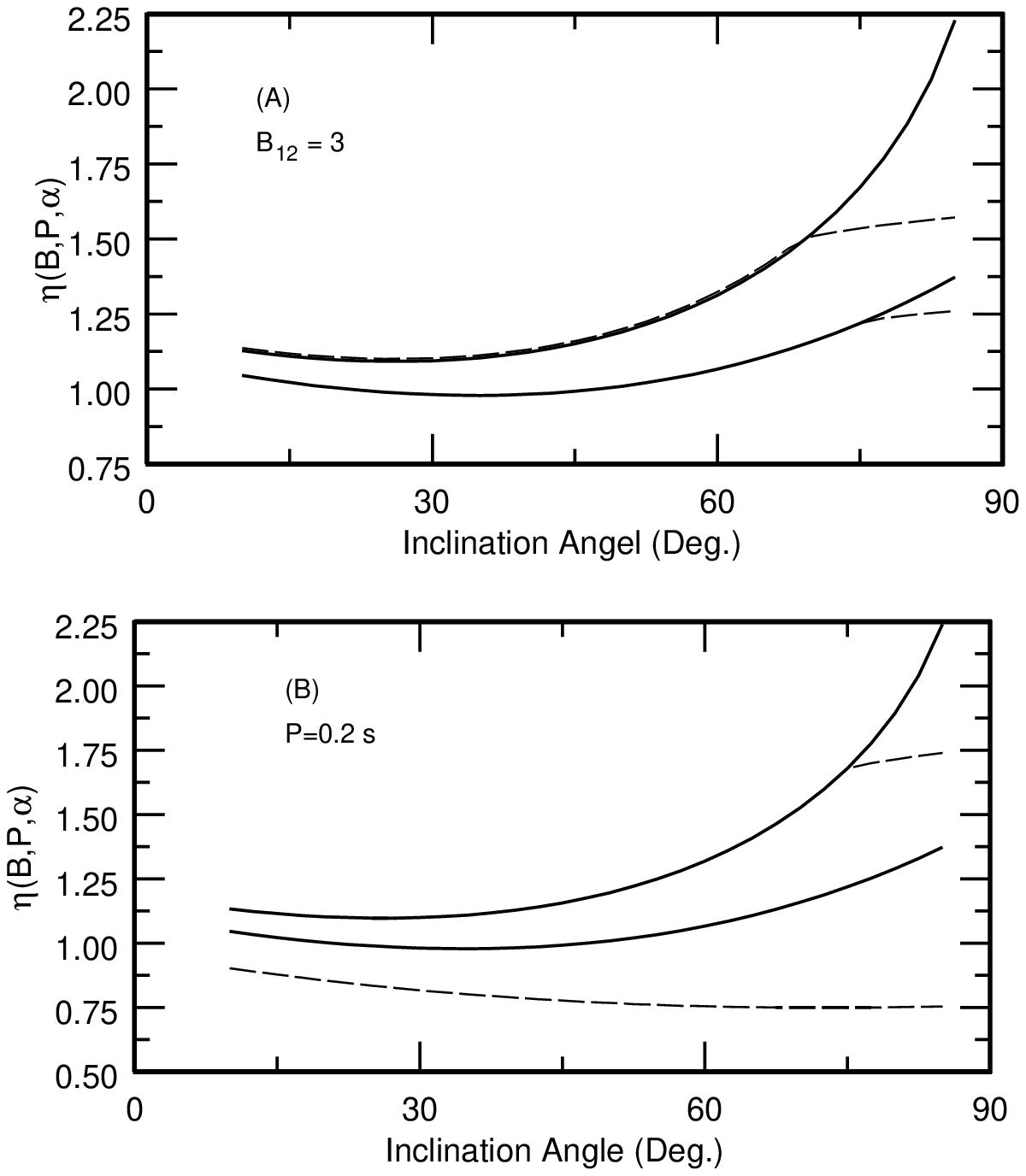}
\caption[newfig1.eps]{Variation of the fractional size of the
outer gap with the magnetic inclination angle for some typical
pulsar parameters. Panel (A): $\eta(P,B,\alpha)$ versus $\alpha$
for a given magnetic field of $3\times 10^{12}$ G. The upper
(lower) solid line is the self-consistent outer gap model (the
cooling X-ray outer gap model) with $P = 0.1s$ and the upper
(lower) dashed line is the self-consistent outer gap model (the
cooling X-ray outer gap model) with $P = 0.3s$ respectively. Panel
(B): $\eta(P,B,\alpha)$ versus $\alpha$ for a given period of 0.2
s. The upper (lower) solid line is the self-consistent outer gap
model (the cooling X-ray outer gap model) with $B_{12} = 3.0$ and
the upper (lower) dashed line is the self-consistent outer gap
model (the cooling X-ray outer gap model) with $B_{12} = 1.5$
respectively. \label{fig1}}
\end{figure}

\begin{figure}
\vbox to2.1in{\rule{0pt}{2.1in}} \includegraphics{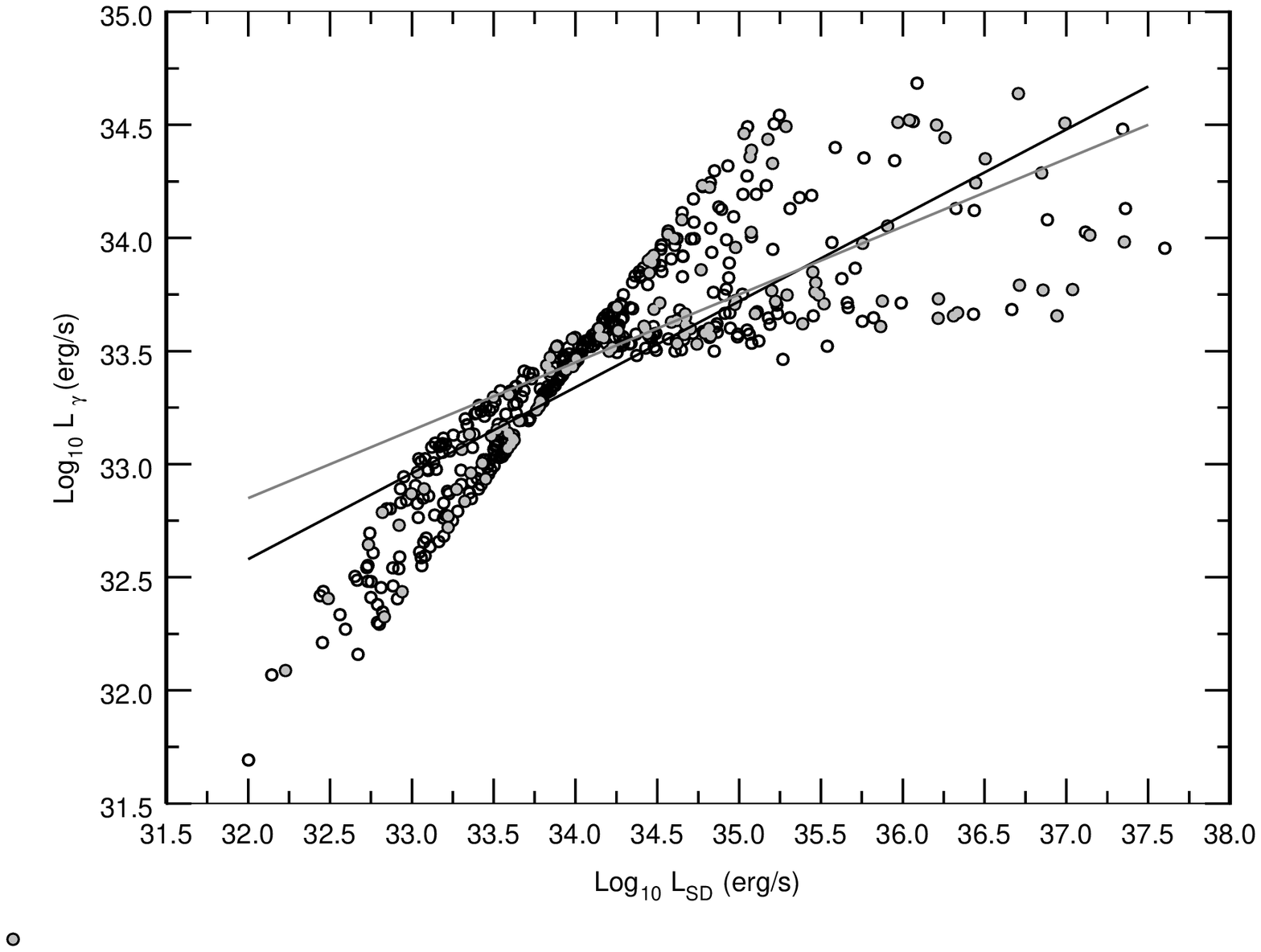}
\caption[newfig3.eps]{The change of $\gamma$-ray luminosity
($L_{\gamma}$ with the spin-down power ($L_{sd}$) in the
$\gamma$-ray pulsar population predicted by our outer gap model.
In our simulation, we have used the EGRET threshold as the minimum
detectable $\gamma$-ray energy flux. Open circles and shaped
circles are the model radio-quiet and radio-loud $\gamma$-ray
pulsars respectively and the solid line is the best fit for all
$\gamma$-ray pulsars with outer gaps. Shaded line is the best fit
for the radio-loud $\gamma$-ray pulsars with outer gaps.
\label{fig2}}
\end{figure}

\begin{figure}
\vbox to3in{\rule{0pt}{3in}} \includegraphics{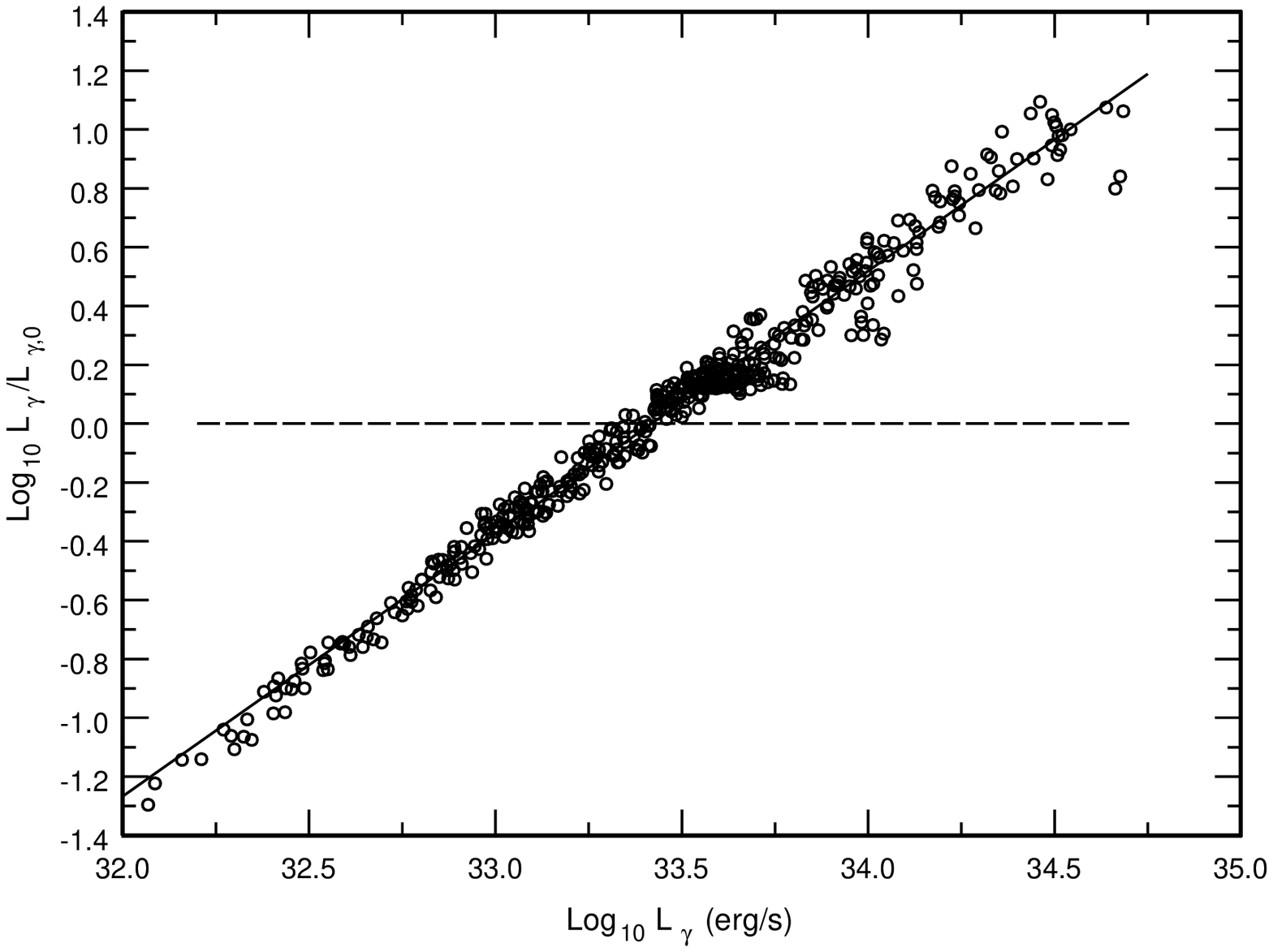}
\caption[newfig4.eps]{$L_{\gamma}/L_{\gamma, 0}$ versus
$L_{\gamma}$ in the $\gamma$-ray pulsar population predicted by
our outer gap model. In our simulation, we have used the EGRET
threshold as the minimum detectable $\gamma$-ray energy flux. Open
circles are the expected data and solid line is the best fit. For
comparison, we show the result given by Zhang \& Cheng (1997) as a
dashed line. \label{fig3}}
\end{figure}

\begin{figure}
\vbox to2in{\rule{0pt}{2in}} \includegraphics{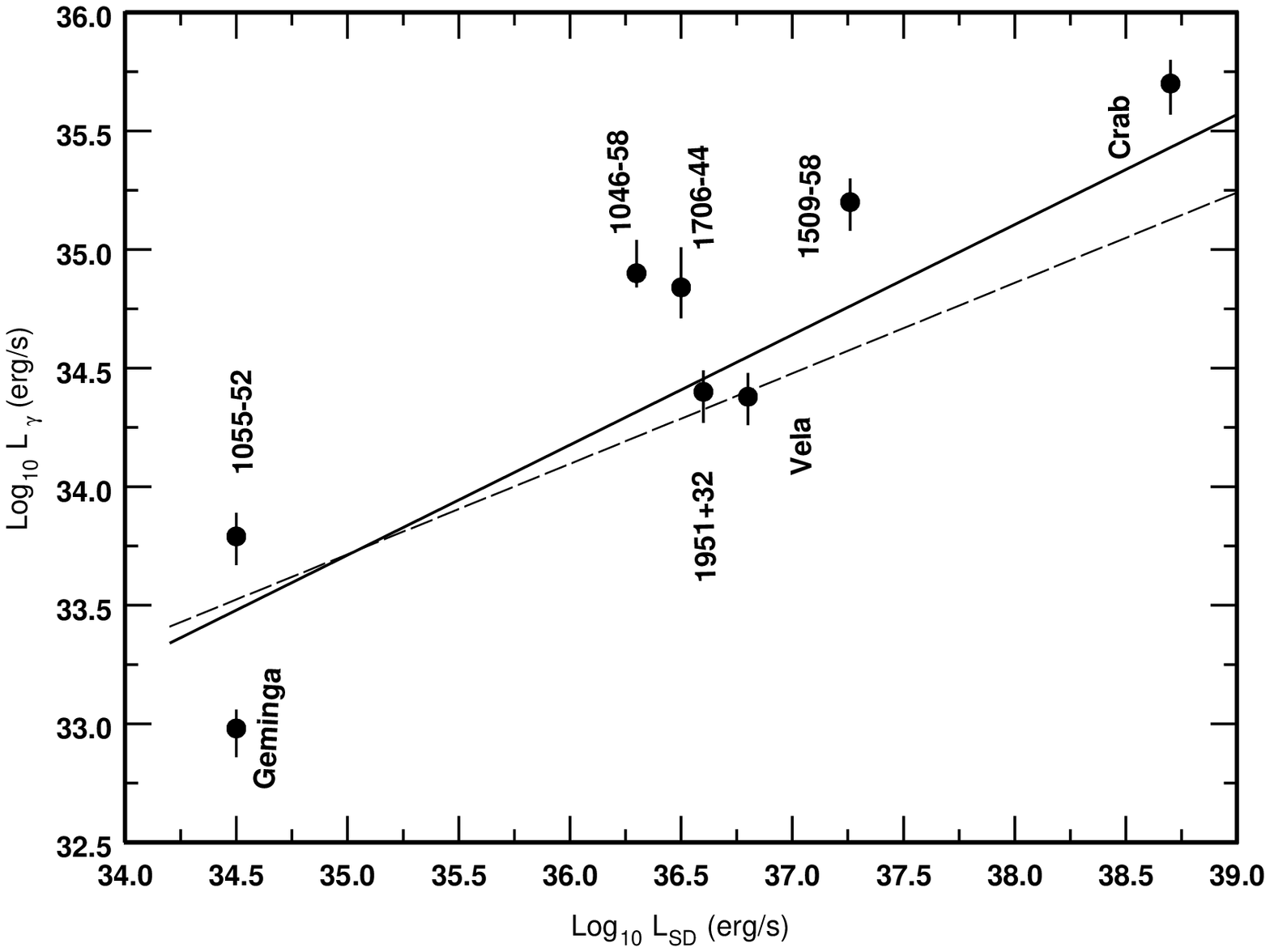}
\caption[newfig5.eps]{$\gamma$-ray luminosity versus the spin-down
power. Solid circles with error bar are the observed data given by
Thompson et al. (2001), solid and dashed lines are our results
given by Eqs. (\ref{LgEGRET}) and (\ref{LgGLAST})
respectively.\label{fig4}}
\end{figure}

\begin{figure}
\vbox to3in{\rule{0pt}{3in}} \includegraphics{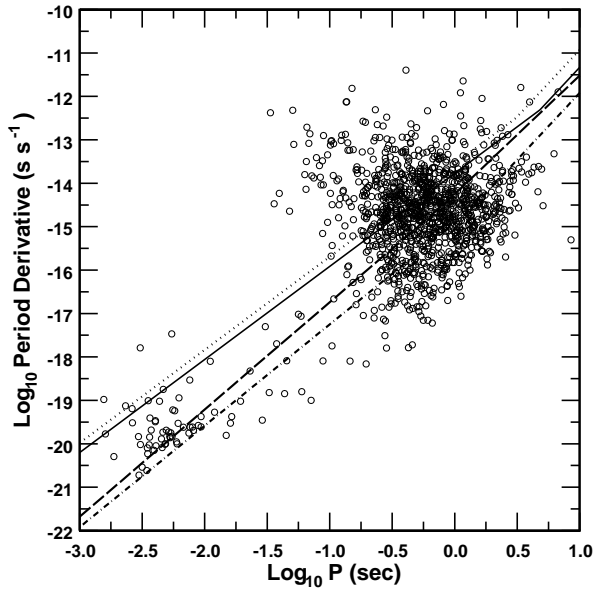}
\caption[newfig6a.eps]{Death lines of the pulsars with outer gaps.
It is assumed that X-rays are produced by the neutron star cooling
and polar cap heating. Two cases of the inclination angle
distributions are considered: (i)an uniform distribution  and (ii)
a cosine distribution. Dotted and solid lines are presented the
death lines given by Eqs. (\ref{deathpcu1}) and
(\ref{deathpccos1}). Dashed and dot-dashed lines represents the
death lines given by Eqs. (\ref{deathpcu2}) and
(\ref{deathpccos2}).  The observed data are taken from see website
http://www.atnf.csiro.au/research/catalogue/) \label{fig5}}
\end{figure}

\begin{figure}
\vbox to3.0in{\rule{0pt}{3.in}} \includegraphics{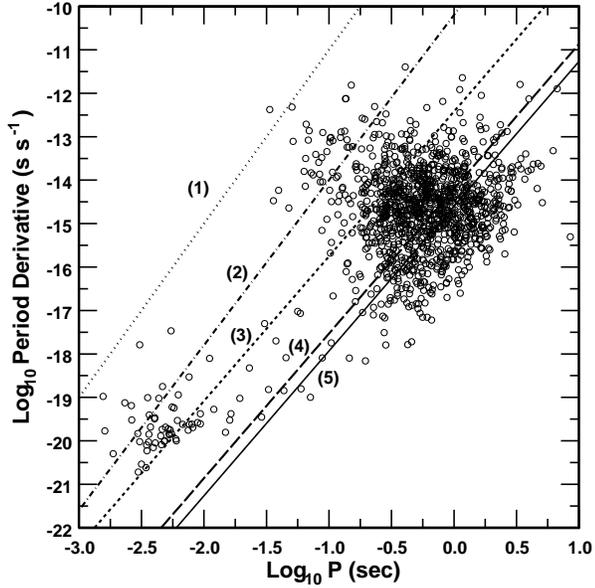}
\caption[newfig6b.eps]{Death lines of the pulsars with
self-sustained outer gaps. The observed data are taken from see
website http://www.atnf.csiro.au/research/catalogue/. Lines (1)
and (2) are given by Eqs. (\ref{death3}) and (\ref{death4})
respectively (Chen \& Ruderman 1993). Line (3) is the death line
(Eq. (\ref{death5})) predicted by Zhang \& Cheng (1997). Lines (4)
and (5) are the death lines of our model for the uniform and the
cosine distributions of the inclination angles, respectively.
\label{fig6}} 
\end{figure}

\end{document}